\documentclass[12pt,a4paper]{article}
\usepackage{graphicx}

\begin{document}
\textwidth=135mm
 \textheight=200mm
\begin{center}
{\bfseries \large Dark Matter Search Experiments}
\vskip 5mm
Wolfgang Rau\vskip 5mm
{\footnotesize {\it Department of Physics, Queen's University Kingston, Ontario, K7L 3N6, Canada}}
\\
\end{center}
{\parbox{135mm}{\scriptsize Prepared for the proceedings of the IVth International Pontecorvo Neutrino Physics School, held in Alushta, Ukraine, from September 26 to October 06, 2010; to be published in {\it Physics of Elementary Particles and Atomic Nuclei} (Pleiades Publishing Inc., Moscow)}}
\vskip 5mm
\centerline{\bf Abstract}
Astronomical and cosmological observations of the past 80 years build solid evidence that atomic matter makes up only a small fraction of the matter in the universe. The dominant fraction does not interact with electromagnetic radiation, does not absorb or emit light and hence is called Dark Matter. So far dark matter has revealed its existence only through gravitational effects. The strongest experimental effort to find other evidence and learn more about the nature of the dark matter particles concentrates around Weakly Interacting Massive Particles which are among the best motivated dark matter candidates. The two main groups of experiments in this field aim for indirect detection through annihilation products and direct detection via interactions with atomic matter respectively. The experimental sensitivity is starting to reach the parameter range which is preferred by theoretical considerations and we can expect to confirm or dismiss some of the most interesting theoretical models in the next few years.
\vskip 10mm

\section{\label{sec:intro}Introduction Evidence for Dark Matter}
The first observation of a significant discrepancy between the amount of matter deduced from optical observations, based on a good knowledge of a typical mass-to-light ratio of galaxies, and the gravitation in the respective system (in this case based on the velocity distribution and the virial theorem) came from Fritz Zwicky's study of the Coma Cluster, published in 1933 \cite{Zwicky}. He concluded that there must be more than two orders of magnitude more matter than could be accounted for by the observed luminous matter. To account for this difference he introduced the idea that there is a vast amount of dark matter present in galaxy clusters.

Not much progress was made towards the understanding of dark matter until Vera Cooper Rubin and coworkers published their observation of rotation curves of spiral galaxies \cite{Rubin}: they documented that the rotational velocities of object outside the visible disk of these galaxies did not follow Kepler's law as expected, but rather stayed constant out to very large radii, implying that galaxies are surrounded by a significant amount of invisible or dark matter. It is worth to note the Jan Hendrik Oort found already in 1932 discrepancies between the observed rotation curve of our own galaxy and the expectation from the luminous matter \cite{Oort}; however, from his observation he was not able to exclude (and actually seemed to favor the interpretation) that this discrepancy may have been caused by an underestimate of luminous matter due to the presence of absorbing matter.

\begin{figure}[ht]
\begin{center}
\includegraphics[width=8cm]{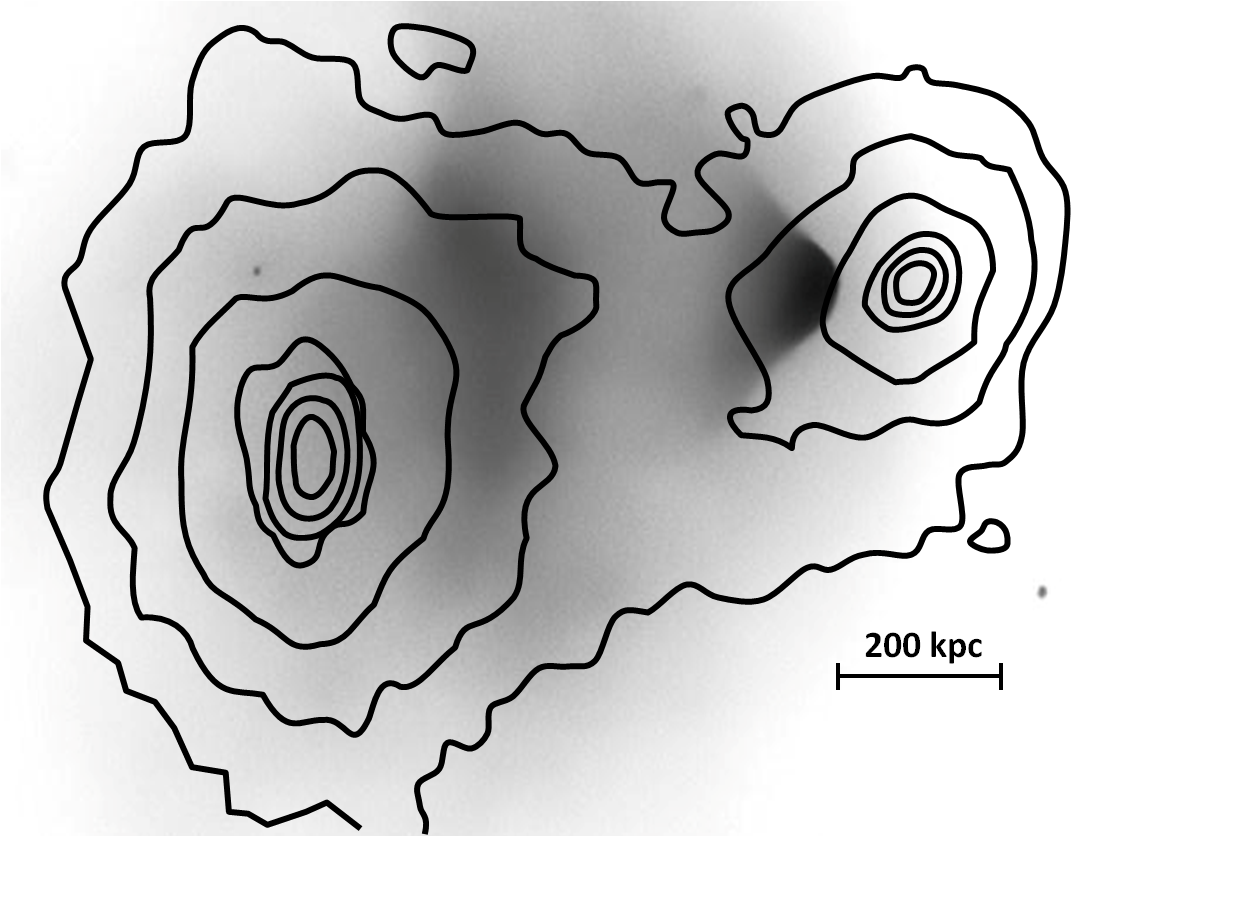}
\end{center}
\vspace*{-0.6cm}
\caption{\small \it X-ray image of the bullet cluster (Credit: NASA/CXC/CfA/ M.Markevitch et al., adapted for scale and color) together with the gravitational potential contours (adapted from \cite{Clowe}). The offset between the gas, indicated by the x-ray emission, and the gravitational potential supports the hypothesis of collisionless dark matter dominating the mass in this system (for more details see text).}
\label{BulletCluster}

\end{figure}

Since then numerous observations have been made which confirm that most of the matter in the universe is dark and non-baryonic (i.e. not consisting of atoms). Only two more shall be mentioned here: the bullet cluster and the cosmic microwave background. The bullet cluster is actually a pair of galaxy clusters, the smaller of which has traversed the larger. The optical emission of the galaxies \cite{Tucker} and the x-ray emission of the hot intergalactic gas \cite{Markevitch} of this structure has been studied and overlaid by the mass distribution contours obtained through the determination of the weak gravitational lensing effect \cite{Clowe}. Due to their small size and large distances, the galaxies behave like collisionless particles while the more or less homogeneously distributed gas collides and is consequently falling behind the galaxies during the encounter. Since the amount of matter in the gas is considerably larger than that in the galaxies one would expect the gravitational potential to follow the gas distribution. However, the center of mass is where the galaxies themselves are found (see figure \ref{BulletCluster}) which supports the hypothesis that most of the matter in this system is collisionless dark matter and disfavors the proposal that the discrepancy between gravitational effects and the amount of luminous matter can be explained by altering the description of gravity or Newton's dynamics (see \cite{Clowe} and references therein).

Finally, the a careful study of the minute temperature fluctuations of the cosmic microwave background leads to the conclusion that more than 80 \% of the matter in the universe does not interact with electromagnetic radiation and as such must be non-baryonic.\cite{WMAP}

\section{Dark Matter Candidates}

The only particle provided by the standard model of particle physics which is stable, uncharged and carries mass, and as such could act as dark matter is the neutrino, but from recent observations we know that the mass of the neutrino is too low to account for the required total mass; and further, the relativistic nature of the neutrino at the time of structure formation in the universe is incompatible with its observed clumpy structure.\cite{Lesgourgues}

If produced in thermal equilibrium in the early universe the dark matter particle needs to be fairly massive to explain the observed large scale structure of the universe. The thermal production also links the dark matter density in the universe to the interaction cross section with ordinary matter which appears to be on the order of the weak scale (see e.g.\ \cite{Jungman}). Thus, a Weakly Interacting Massive Particle (WIMP) would be a prime candidate to solve the dark matter problem ('weak' in this case is generic and not necessarily the Weak Interaction mediated by the exchange of W or Z bosons). 

Interestingly, several extensions of the standard model of particle physics proposed for completely independent reasons predict new particles that would appear as WIMPs, most notably Supersymmetry with the neutralino, a mixture of the supersymmetric partners of the uncharged bosons (see e.g.\ \cite{Bertone}), or universal extra dimensions with the Kaluza-Klein particles \cite{Hooper}.

While certainly the most discussed candidate, it should be noted that a WIMP is not the only potential solution to the dark matter problem. Axions, originally introduced to solve the problem that no CP-violation is observed in the strong interaction, could be produced non-thermally in a phase transition in the early universe in considerable quantities and, depending on their exact properties, act as dark matter even if their mass is much smaller than that of other known elementary particles (for a review see e.g.\ \cite{Raffelt}). There are several past and present experiments searching for axions; however, only one, the Axion Dark Matter Experiment (ADMX), is sensitive dark matter axions.\cite{ADMX} This experiment uses a tunable radio-frequency (RF) cavity operated at low temperature (1.6 K, to reduce the noise) in a strong magnetic field (7.9 T). Through their scattering on the virtual photons of the magnetic field, the axions can be converted into real photons. If the cavity frequency matches the axion mass this conversion would be resonantly enhanced and the measured power in the cavity would increase. No signal has been observed so far which excludes light axions ($\sim$2 $\mu$eV) as dominant dark matter component in certain axion models. To test the full cosmologically relevant parameter range and other axion models, the sensitivity needs to be increased by about an order of magnitude and the mass range needs to be extended. Improvements of the experiment are underway.

Even in a thermal production scenario the above mentioned link between production and interaction cross section can be broken, leading to potentially very weakly interacting dark matter particles, such as gravitinos (the supersymmetric partner of the graviton). The respective theories are less appealing though due to the added mechanisms to break said link, and not the least because the emerging particles are not accessible to experimental dark matter searches.

\section{Indirect Detection of WIMPs}

Supersymmetric WIMPs are Majorana particles so that we would expect WIMP-WIMP annihilation in regions of high WIMP density. Such regions are e.g.\ the centers of galaxies, but also the core of astronomical objects such as sun and earth, where WIMPs are expected to have accumulated over their live time through scattering processes. The detection of annihilation products of WIMPs is usually referred to as indirect dark matter detection. Even if WIMPs are Dirac particles annihilation is possible if the densities of particles and anti-particles are comparable.

If the annihilation happens inside a dense body, the only products that can be detected are high energy neutrinos. Large neutrino detectors such as SuperKamiokande \cite{SK} and IceCube \cite{IceCube} have searched for such signals but so far only upper limits on the high energy neutrino flux from the center of the sun or the earth can be given.

If the annihilation occurs in free space, other types of radiation can be detected. To distinguish a potential dark matter signal from cosmic radiation from more conventional sources, specific characteristics need to be found. If WIMPs annihilate directly into gamma rays, one would expect a line in the observed spectrum at the WIMP mass; if gammas emerge together with other particles, we only would expect some enhancement of the spectrum below the WIMP mass, but in any case the respective signal should originate a places of higher dark matter density. VERITAS \cite{VERITAS} is a ground based Cherenkov telescope for gamma ray astronomy and has among other things searched for an enhancement of gamma rays from the center of neighboring dwarf galaxies without a positive signal so far. Data from the EGRET satellite have been interpreted as evidence for dark matter \cite{EGRET}, but this interpretation is incompatible with data from the more recently launched Fermi satellite \cite{Fermi_g}. 
Anti-matter is a very rare product of conventional sources of cosmic radiation and has therefore been proposed as possible indicator for dark matter annihilation. PAMELA, a satellite based instrument specifically designed to search for anti-particles in cosmic radiation, has reported an enhancement of positrons around 100~GeV \cite{PAMELA}. If originating from dark matter annihilation a respective enhancement of anti-protons would be expected but is not observed. Special dark matter models could avoid hadron production, but more conventional explanations exist as well, such as nearby pulsars. 
ATIC is a balloon borne experiment which can discriminate between leptons and hadrons, but due to the lack of a magnetic field not between particles and anti-particles. Here a peak is observed in the electrons/positron spectrum in the few hundred GeV range, which also has been interpreted as possible evidence for dark matter annihilation \cite{ATIC}, however other experiments originally designed for gamma ray detection like the ground based H.E.S.S. \cite{HESS} and the LAT at the Fermi satellite \cite{Fermi_b} did not see this peak and only show a slight excess above conventional model calculations. Several authors have attempted to find a common dark matter interpretation of the PAMELA, ATIC, HESS and Fermi data (see e.g.\ \cite{PamelaAticHess_I}, \cite{PamelaAticHess_II}), however the observed rate is too high by several orders of magnitude to be compatible with the observed relic density of dark matter given the average density of dark matter in our galaxy. Non-standard astrophysics and particle physics enhancement mechanisms have to be invoked such as strong local dark matter overdensities and Sommerfeld enhancement, and even then it is difficult to justify this interpretation.

\section{Direct Detection of WIMPs}
\begin{figure}[ht]
\begin{center}
\includegraphics[width=12cm]{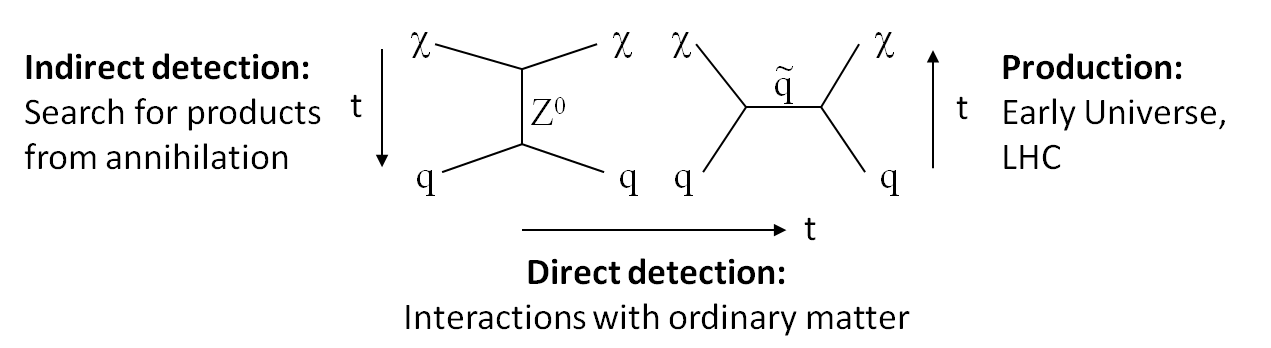}
\end{center}
\vspace*{-0.6cm}
\caption{\small \it The two Feynman diagrams above describe two possible processes for the production of Supersymmetric WIMPs in the early universe as well their direct or indirect detection, depending on which direction we chose for the time axis. Since the relic density is correlated with the production cross section, we also expect a link between the relic density and the detection probability.}
\label{Feynman_DM}

\end{figure}

If WIMPs are produced in thermal equilibrium in the early universe or annihilate into standard model particles, we expect that they also directly interact with ordinary matter, since the Feynman diagrams describing these processes are essentially the same (see figure \ref{Feynman_DM}). WIMPs interacting in terrestrial detectors would primarily be those gravitationally bound to our galaxy. Since the escape velocity is a few hundred km/s \cite{HaloParam} we can easily estimate that the maximum energy transfer from a WIMP to an electron initially at rest is at most in the eV range, while the energy transfer to an atomic nucleus would typically be in the range of tens of keV. Therefore direct detection experiments typically search for nuclear recoils induced by WIMPs.

It turns out that in this range of momentum transfer the scattering amplitudes of all nucleons in a given nucleus interfere constructively such that the scattering amplitude is proportional to the nucleon number A. Since the probability is proportional to the square of the amplitude we expect a cross section proportional to A$^2$ which favors heavy target nuclei. This holds for medium size nuclei; for very heavy nuclei and higher momentum transfer we lose coherence and the effective cross section increases slower with A. 

Depending on the underlying theory however, the WIMP may couple primarily to the spin of the nucleons. In this case the coherence is a disadvantage, since the scattering amplitudes of nucleons with opposite spin cancel out. For spin-dependent interaction the favored target nuclei are consequently those with an unpaired nucleon and a high spin factor

Within a given theoretical framework the interaction rate can be calculated. Minimal Supersymmetric models are among the most popular extensions of the standard model of particle physics, but even within this particular framework the predicted WIMP-nucleon interaction cross section spans many orders of magnitude. Typical values for the spin-independent cross section are between 10$^{-6}$~pb and 10$^{-11}$~pb \cite{Austri}. Such small cross sections imply that large target masses and long measurement times are required; at the lower end of the cross section range typical interaction rates are a few per ton per year. 

These low expected rates pose a major challenge considering that typical background rates from environmental radioactivity and cosmic radiation are much higher. As a protection against cosmic radiation dark matter search experiments are usually installed deep underground. Worldwide a large number of underground laboratories exist, many of which house present and/or future dark matter experiment (see table \ref{Labs} for a selection). Environmental radioactivity is mediated by a combination of active and passive shielding. Typical shielding materials for gamma radiation are copper and lead, while the neutron flux is usually reduced by water, polyethylene or paraffin. Active muon veto detectors (plastic or liquid scintillators) are used to identify events that are induced by the remaining cosmic ray muon flux. The internal radioactivity is reduced by a careful selection of construction materials, aiming for the lowest possible concentration of radioactive trace contaminants.

\begin{table}
		\footnotesize
		\begin{center}
		\begin{tabular}{lll}
		\hline
		\hline
		Laboratory				&Depth	&Experiments\\
											&[mwe]	& \\ \hline
	  \ \\[-2ex]
		Soudan MN, USA		&2000		&CDMS/SuperCDMS, CoGeNT\\
		Yangyang, Korea		&2000   &KIMS\\ 
		Canfranc, Spain		&2500		&IGEX\\
		Kamioka, Japan		&2700		&XMASS, SuperKamiokande\\
		Bulby, UK					&3200		&ZEPLIN\\
		Gran Sasso, Italy	&3500		&DAMA/LIBRA, CRESST, WARP\\
		Homestake, ND, USA
		     \hspace*{2em}&4500
		             \hspace*{2em}&LUX\\
		Modane, France		&4800		&EDELWEISS\\
		SNOLAB, Canada		&6000		&PICASSO, DEAP/CLEAN\\ 
		\hline
		\hline
		\end{tabular}
		\end{center}
		\vspace*{-0.6cm}
\caption{\small \it Some of the underground laboratories which house dark matter experiments; the approximate effective shielding depth is measured in meter water equivalent (mwe). Only experiments that are discussed in this article are listed}
\label{Labs}
\end{table}

The above measures greatly reduce the background, but there is still an appreciable background rate left, which would limit the sensitivity. Therefore most experiments employ some form of background discrimination. This is typically based on the fact that most background radiation is ionizing radiation which interacts with the electrons, while the WIMPs are expected to reveal themselves through nuclear interactions as discussed above. 

\begin{figure}[ht]
\begin{center}
\includegraphics[width=8cm]{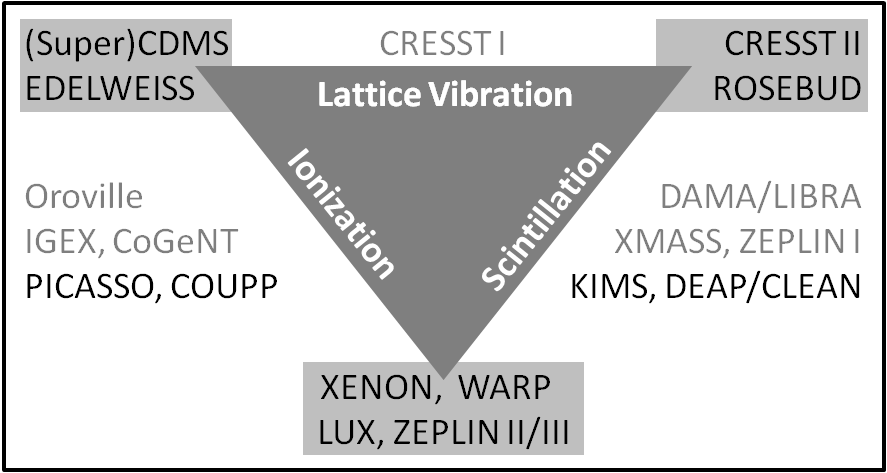}
\end{center}
\vspace*{-0.6cm}
\caption{\small \it Some examples of dark matter experiments and the technologies they employ. The experiments at the corners of the triangle achieve a high event-by-event discrimination efficiency due to the combination of two signals. But in some cases also a single signal provides good discrimination power due to threshold effects (PICASSO and COUPP) or pulse shape discrimination (KIMS and DEAP/CLEAN).}
\label{DM_Experiments}
\end{figure}

Three detection principles are the basis of most particle detectors: the ionizing effect of a particle interaction, scintillation light from electronic excitation or a thermal signal from lattice vibrations. The characteristics of these three signal types often differ for nuclear and electron recoils giving a handle on discriminating the remaining background. Figure \ref{DM_Experiments} shows some examples of experiments and their detection technologies. Combinations of two signals provide very efficient electron recoil background discrimination, but in some cases (PICASSO, COUPP, KIMS and DEAP/CLEAN among the given examples) also a single signal can give good discrimination (for more details see the experiment descriptions below).

In ionization (and also scintillation) detectors, the signal size of nuclear and electron recoils of the same energy is different. This {\it quenching} effect can be described by the {\it quenching factor}, QF, which is just the ratio of the signal size for the two types of events {\footnote{Different conventions are used; we adopt here the definition  QF\ :=\ NR/ER, where NR and ER are, respectively, the signal sizes of nuclear and electron recoils of the same energy}}. The quenching factor is usually a function of the energy. The energy scale is mostly determined with gamma calibration sources and therefore given in electron equivalent units (i.e.\ the energy that a gamma would have to have to produce a signal of the observed size), usually expressed as keVee. To find the nuclear recoil energy scale, the electron equivalent scale has to be divided by the quenching factor.

In the following subsections we will discuss some of the past, present and future direct dark matter search experiments in more detail. The list is certainly not exhaustive but the attempt has been made to represent the major technologies and include some of the most important experiments. The main results will be discussed at the end.

\subsection{Ionization Detectors}
Among the first detectors used for dark matter search were germanium detectors designed to search for neutrinoless double beta decay. Technologically, the detectors are standard high purity germanium detectors operated at liquid nitrogen temperature. The expected rate for this process is also very low and thus the detectors have to fulfill strict background requirements. 

One of the earliest experiments in this category was a double beta experiment located in a relatively shallow laboratory (600~mwe) at a water power plant in Oroville, reporting limits on some early dark matter models already in 1988. Data have been taken with a 0.9~kg Ge detector, surrounded by a NaI anti-compton detector, borated polyethylene (PE) to reduce the neutron flux and 20~cm of lead. A threshold of 3~keVee and a background count rate of $\sim$0.5~/keV/kg/day lead to a sensitivity for the WIMP-Ge cross section in the 10$^2$~pb range, which corresponds to a spin-independent WIMP-nucleon cross section in the 10$^{-2}$~pb range for WIMPs in the mass range of tens of GeV/c$^2$.\cite{Oroville} 

The IGEX experiment located at the Canfranc laboratory in Spain, also searching for neutrinoless double beta decay, optimized the shielding (25 cm of archaeological Pb with very low $^{210}$Pb content and 20~cm of 70 year-old low activity lead, a muon veto counter and a 40~cm neutron moderator out of borated water and polyethylene) and the background of one of their Ge detectors (2.2~kg) at low energy. The achieved threshold is 4~keV and the background rate is 0.04~/keV/kg/day above 20~keV, increasing up to 0.2~/keV/kg/day near threshold. Results published in 2002 showed a sensitivity around 10$^{-5}$~pb.\cite{IGEX} 

Recently the attempt has been made by the CoGeNT collaboration to optimize the germanium detector technology for low energies and low background. One of the electrodes of the employed detector has a rather small area ({\it point-contact}) which minimizes the capacity and thus the electronic noise and the threshold. The detector is surrounded by a NaI anti-compton detector, lead and neutron moderator and two layers of active muon veto. Even though the detector has a relatively low mass (440 g, three quarter of which are active), the low threshold (0.4~keVee) and decent background levels ($\sim$1~/keV/kg/day around 3~keV) helped to reach a sensitivity in the same cross section range as IGEX, albeit for WIMPs with considerably lower mass (few GeV/c$^2$).\cite{CoGeNT} 

\subsection{Room Temperature Scintillators}
The largest operating experiment for direct dark matter search is DAMA. The target consists of large ($\sim$10~kg) scintillating NaI crystals read out by photo-multiplier tubes (PMTs) at room temperature in a well shielded and controlled environment. NaI does not provide a strong event-by-event discrimination of electron recoil background. Therefore DAMA follows a unique strategy to still get a handle on the background: the sun orbits our galaxy with a velocity of $\sim$220~km/s, while the earth rotates around the sun with $\sim$30~km/s. Assuming the dark matter halo around our galaxy has no net angular momentum, the relative velocity between the WIMPs and the detector changes over the course of the year. For a given energy threshold this would lead to an annual modulation of the interaction rate with a known phase. The target is surrounded by high purity copper and lead against gamma radiation and paraffin and polyethylene to moderate neutrons.
The DAMA/NaI project operated $\sim$100~kg of NaI for seven annual cycles and was followed by the DAMA/LIBRA project with a target mass of $\sim$250~kg, operated for 6 annual cycles and achieved a background rate of $\sim$1~/keV/kg/day. The total amount of data accumulated is more than 1~ton-year.\cite{DAMA_I}\cite{DAMA_II} 

KIMS is a more recent experiment, but with a significant mass as well: 3.4~ton-days have been collected with four large (8.7~kg) scintillating CsI crystals, also read out by PMTs, surrounded by a composite shielding with copper (10~cm), polyethylene (10~cm), lead (15~cm) and a 30~cm liquid scintillator veto. Based on differences in the average pulse shape between nuclear and electron recoil events, a background rate of order of 0.2~/keV/kg/day has been achieved down to the threshold of 3~keVee, leading to a sensitivity in the 10$^{-6}$~pb range for the spin-independent WIMP-nucleon cross section. However, given the target material, this experiment has also a good sensitivity to spin-dependent interaction.\cite{KIMS}

\subsection{Cryogenic Detectors}
The CDMS experiment was the first to develop a detector technology which allows for a highly efficient event-by-event discrimination between electron and nuclear recoil events based on the recording of two independent signals. The detectors are based on germanium or silicon single crystals with a diameter of 7.5~cm (3~inch) and a thickness of 1~cm ($\sim$240~g for Ge and 100~g for Si). The energy of the interaction is determined via the induced lattice vibrations ({\it phonon signal}) while the ionization signal distinguishes between event types based on the quenching effect. The phonon signal, which is essentially independent of the interaction type and thus provides a reliable energy measurement, is measured with a thin superconducting tungsten film evaporated onto one face of the crystal. The operating temperature of the detectors is roughly 40~mK and the phonon sensor is held in the transition between superconducting and normal state leading to a strong dependence of the resistance on the temperature. The sensor is structured photo-lithographically into a large number of small filaments wired in parallel; aluminum fins are attached to the tungsten filaments to enlarge the effective sensor area without increasing the heat capacity (which would reduce the sensitivity). The large area of the sensor allows the collection of the phonon energy long before thermalization, thus preserving additional information about the event, specifically information about the location. The charge carriers are collected with a low voltage (typically 3~V) applied to an aluminum electrode on the back side of the crystal. 

The ratio of the two signals gives a very efficient discrimination between electron and nuclear recoils for events with more than about 10~keV recoil energy, happening in the bulk of the detector, but unfortunately events close to the detector surface suffer from a reduced charge signal, which moves some of the electron recoil background into the signal region. The position sensitivity of the detector helps to identify and remove surface event background. 

CDMS operated 30 detectors (19 Ge and 11 Si) between 2006 and 2008 and collected a total of more than 300~kg-days (after all analysis cuts) with only a few background events. With an event rate in the signal region of only $\sim$10$^{-4}$~/keV/kg/day above the analysis threshold of 10~keV this experiment provided the best sensitivity for the spin-independent WIMP-nucleon cross section for most of the last decade.\cite{CDMS}

SuperCDMS continues to improve the CDMS detector technology, with larger detector modules and an even better background discrimination. After the present detector R\&D phase is concluded, dark matter search will continue in 2011 with a total target mass of 10-15~kg. To overcome the expected limitation by cosmic radiation at the end of this phase, SuperCDMS plans to move to SNOLAB and build a new setup with a total target mass in the 100~kg range.

EDELWEISS uses the same basic idea as CDMS. However, the thermal signal is measured with a small neutron-transmutation-doped thermistor (NTD). This makes detector production much easier, but does not provide the additional discrimination of surface events. The experiment collected a total of 62~kg-days with their first generation detectors which achieved a background rate in the the signal region of $\sim$0.03~/keV/kg/day between 15~keV and 30~keV but about a factor of 100 lower at higher energy.\cite{EDELWEISS_I}

A recent development with a new electrode structure provides surface event rejection based on the ionization signal and leads to a larger signal acceptance. Ten 400~g detectors with this new technology have been used to collect a total of 322~kg-days of WIMP search data with an energy threshold of 20~keV. The nominal background rate is very similar to that achieved by CDMS, albeit with a somewhat higher threshold.\cite{EDELWEISS_II}

The CRESST experiment started with cryogenic sapphire detectors with no active background discrimination \cite{CRESST_I}, but then developed scintillating cryogenic detectors based on 300~g CaWO$_4$ single crystals. The phonon signal is detected with a superconducting tungsten sensor like in CDMS, but with a much simpler design. The light is detected with a separate low-mass cryogenic detector consisting of a thin silicon coated sapphire disk, equipped as well with a tungsten sensor. These detectors have an extremely low energy threshold of $\sim$20~eV, necessary to detect the faint scintillation signal. Each detector module consists of a scintillating crystal and a light detector surrounded by a reflective housing to maximize the light collection efficiency. 

In CaWO$_4$ the quenching is not only different for electron and nuclear recoils but also differers for the different types of nuclei in the crystal. This gives an additional diagnostic tool to partially separate neutron background (WIMPs prefer the heavy tungsten as scattering partner due to the A$^2$ dependence of the cross section, while neutrons can transfer energy more efficiently to oxygen nuclei), identify background sources or study a potential signal under different hypothesis.

CRESST has published data from a first run with an improved setup (a muon-veto and a 45~cm polyethylene neutron shield was installed in addition to previously existing gamma shield of 14~cm of copper and 20~cm of lead) where a background rate of order of 10$^{-3}$~/keV/kg/day above the 10~keV threshold was achieved in a data set representing 48~kg-days of exposure.\cite{CRESST_II}

\subsection{Noble Liquids}
Noble elements are very good scintillators and since they do not engage in chemical reactions they can be purified very well. Those are ideal conditions to use them as detector material for dark matter searches where low energy events need to be detected at a very low rate. A slight drawback is that the determination of the energy scale requires a good knowledge of the quenching at low recoil energies, which is so far not well understood, since measurements in this regime are very challenging.

ZEPLIN is the name for a series of projects using liquid xenon to search for dark matter. The first detector, ZEPLIN I, used a total amount of 5~kg of xenon in a teflon lined copper vessel with a fiducial volume of 3.2~kg, watched by three PMTs. The detector was surrounded ($\sim3\pi$) by a 30~cm thick liquid scintillator veto and a 25~cm lead shield ($4\pi$). A total of $\sim$300~kg-days of data was selected and the attempt was made to discriminate electron recoil background based on the pulse shape. However, no convincing data on the pulse-shape discrimination could be presented due to an instrumental failure before the calibration measurements could be completed. The background rate above a threshold of roughly 3~keVee (the quenching factor was not known precisely but was assumed to be 0.22, so the threshold would correspond to slightly above 10~keV nuclear recoil energy) was roughly 10~/keV/kg/day before, and of order of 10$^{-1}$~/keV/kg/day after the controversial pulse shape discrimination.\cite{ZEPLIN_I}

The background discrimination in ZEPLIN II and ZEPLIN III is based on the measurement of an ionization signal in addition to the scintillation light: since noble elements are chemically inert, free electrons have a long life-time and can be drifted over considerable distances. These electrons are then extracted from the liquid. In the gas phase they are accelerated by a set of wires at positive high voltage and thus produce a secondary scintillation light. In such two-phase detectors the events can be localized in the z-direction by the time difference between the first (S1) and the second (S2) scintillation signal given the known drift velocity of the electrons, and in the the x- and y-direction by the position of S2, thus allowing for an efficient fiducialization.

The geometry of the ZEPLIN III detector is flat (a disc of roughly 27~cm in diameter and a thickness of 3.5~cm) with a fiducial volume of 6.5~kg.  The experiment uses 31 PMTs immersed in the liquid. The quenching factor depends on the electric field (a higher field pulls more electrons away from the initial interaction region, reducing the recombination probability and thus the scintillation efficiency). To determine the recoil energy scale a zero-field quenching factor of 0.19 is assumed and the respective correction for the applied electric field is applied. The event rate in the effective exposure of $\sim$130~kg-day was roughly  $5.6\times 10^{-3}$~/keV/kg/day in the energy range used for analysis (5-15~keVee).\cite{ZEPLIN_III}

The XENON10 experiment uses the same basic technology, but with PMTs in the gas phase as well as in the liquid phase. The active volume is defined by a teflon cylinder with a diameter of 20~cm and a height of 15~cm and the fiducial volume contains 5.4~kg of xenon. The active volume is surrounded by an additional 10~kg of liquid xenon, a steel cryostat with a total mass of $\sim$180~kg and 20~cm of each PE and Pb. The setup was operated for 58.6 live-days; taking into account the nuclear recoil acceptance of about 50\% the effective exposure is $\sim$160~kg-days. The nuclear recoil energy scale is determined in exactly the same way as discussed above for ZEPLIN III. The deduced nuclear recoil energy threshold is 4.5~keV and the effective event rate in region of interest (4.5-29.6~keV) after all cuts is around $2\times 10^{-3}$~/keV/kg/day.\cite{XENON10}

XENON100 is one of the successor projects of XENON10. The experimental setup has been enlarged and more care has been taken to reduce the background. The inner active xenon volume of roughly 30~cm by 30~cm is surrounded by an active xenon-veto detector, and a composite shield of copper (5~cm), PE (20~cm), Pb (20~cm) and an additional partial outer neutron moderator (PE and water). A first data set with a 40~kg fiducial volume taken during the initial 11.7~days of operation in low-background mode has been analyzed and yielded a background rate of $2\times 10^{-3}$~/keV/kg/day before electron recoil discrimination and no nuclear recoil event left in an effective exposure of 161~kg-days. The very low rate before discrimination is due on the one hand to the high purity and very efficient self-shielding of the xenon thanks to its high density and high atomic number which makes it an ideal gamma shield, and on the other hand the good position resolution (a few mm) which makes the fiducialization straight forward and very effective.

LUX \cite{LUX} is a second project following the original XENON10 with a slightly larger target than XENON100. Unfortunately the deployment at the new Sanford Underground laboratory, being constructed at the old Homestake laboratory site, had to be postponed several times since the laboratory is not ready yet.

XMASS is a Japanese project with a total of about 800~kg of liquid xenon in a spherical vessel observed by a large number of PMTs. This is a single phase detector with no, or only very limited discrimination against electron recoil background. However, as we have seen already in XENON100, the self-shielding in xenon is very effective, so if the fiducial volume can be defined with good precision, the final background rate may be very low, if the xenon can be purified well enough with respect to radioactive noble gas species such as $^{85}$Kr and $^{222}$Rn.

Argon also has been proposed as dark matter target; its lower atomic mass disfavors it compared to xenon, but the scintillation efficiency is very high, and in contrast to xenon, it provides a very effective background discrimination through the pulse shape. The problem with argon however is that if extracted from the atmosphere, it contains the radioactive isotope $^{39}$Ar with an activity of about 1~Bq/kg. This leads to a high intrinsic background rate and extreme requirements with respect to background discrimination.

WARP has built a 2-phase prototype liquid argon detector with a fiducial mass of 1.8~kg which achieved a good discrimination of the $^{39}$Ar background and reached a background rate of $2\times 10^{-3}$~/keV/kg/day after discrimination in the 20-40~keVee range. \cite{WARP}
A major uncertainty here is once more the nuclear recoil energy scale. The WARP collaboration has produced a large (order of 100~kg) detector at Gran Sasso with a massive active liquid argon shield, however unforeseen technical difficulties seem to prohibit a timely start of the experiment.

While WARP is a two phase detector it seems that the major part of the discrimination power comes from the pulse shape. Based on this, a single phase liquid argon project has been proposed \cite{DEAP} and two detectors, DEAP 3600 and MiniClean with fiducial masses of 1000~kg and 100~kg respectivcely are under construction at SNOLAB. The pulse-shape based discrimination is expected to provide a sufficient electron recoil background reduction to completely suppress the intrinsic radioactivity, but it is also being considered to fill the detector with argon extracted from underground sources which are depleted in $^{39}$Ar by a factor of 20 or more.\cite{DepletedAr}

\subsection{Superheated Liquids}
Superheated liquids have been used in particle detectors early on in the form of bubble chambers. 
The PICASSO project is based on this technology, but instead of a monolithic bubble chamber, the PICASSO detectors consist of tiny droplets immersed in a gel matrix. This reduces the spontaneous nucleation rate which are typically observed at surfaces and also enable a continuous operation: a particle interaction will trigger the evaporation of a single droplet if the ionization density from the event is high enough. The pressure wave created by the evaporation is picked up by an array of piezo sensors. Such a detector can typically be operated for a day before the droplets are re-condensed by applying elevated pressure. The operating conditions are chosen such that gamma interactions cannot trigger a phase transition due to their relatively low ionization density, while nuclear recoils (and unfortunately also alpha particles) can be detected. The advantage of this technology are the relatively low costs and simple detector production, while a clear disadvantage is that there is no energy information available. PICASSO has operated several of these detectors with a total volume of 4.5~l, containing of order of 70~g of the main WIMP target fluorine each. With this target the experiment is mainly sensitive to spin-dependent WIMP-nucleon interactions. The background rate is of order of 0.01~/g/h for a nuclear recoil threshold of only 2~keV. Even though this background seems high, PICASSO reached a very competitive sensitivity for the proton spin-dependent WIMP-nucleon cross section with an exposure of 14~kg-days \cite{PICASSO}. Further background reduction can be expected with the recent discovery of a way to discriminate alpha background based on the pulse shape of the events.\cite{PICASSO_alpha}

The COUPP project is based on the same idea of particle detection, however using a monolithic bubble chamber. The advantage is less inactive material in the detector, but on the other hand, each event requires the whole detector to be re-compressed for a while leading to a significant dead time. This requires an extremely good control of the surface to avoid spontaneous nucleations. The target contains both fluorine and iodine leading to a good sensitivity for both, spin-dependent and spin-independent WIMP-nucleon scattering. COUPP has produced several chambers of different size. Data have been published from a 1.5~kg chamber (52~kg-days, \cite{COUPP_I}) and a 3.5~kg chamber (28.1~kg-days, \cite{COUPP_II}). 

\subsection{Results}
Most experiments have not seen any evidence of interactions of WIMPs. In such cases the result is represented as an upper limit on the WIMP-nucleon cross section as function of the WIMP mass. The calculation of such a limit requires that certain assumptions are made regarding the astrophysical WIMP properties. Unfortunately not all of these parameters are very well known and different experiment chose slightly different values; however the differences are in most cases not very significant so that we still can usefully compare the results from the different experiments.

\begin{figure}[ht]
\begin{center}
\includegraphics[width=10cm]{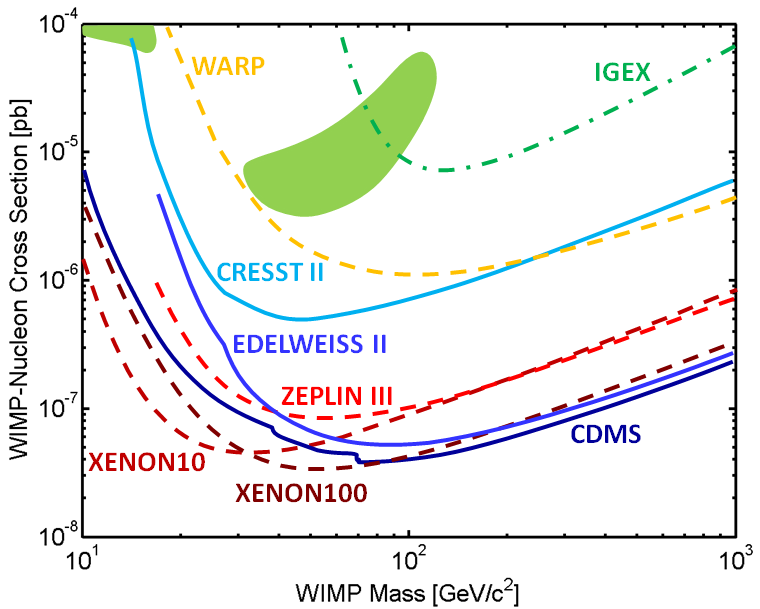}
\end{center}
\vspace*{-0.6cm}
\caption{\small \it Limits on the spin-independent WIMP-nucleon cross section from some of the experiments discussed here. From top to bottom ordered by the minimum of the curves: IGEX \cite{IGEX}, dash-dotted; WARP \cite{WARP}, light (orange) dashed; CRESST II \cite{CRESST_II}, light solid; ZEPLIN III \cite{ZEPLIN_III} medium light (light red) dashed; EDELWEISS II \cite{EDELWEISS_II}, medium solid; XENON10 \cite{XENON10}, medium dark (medium red) dashed; CDMS \cite{CDMS}, dark solid; XENON100 \cite{XENON100}, dark (dark red) dashed. Also shown is the 5$\sigma$ region allowed by DAMA as interpreted by Savage et al.\ \cite{Savage}, shaded region.}
\label{SI}
\end{figure}

\begin{figure}[ht]
\begin{center}
\includegraphics[width=10cm]{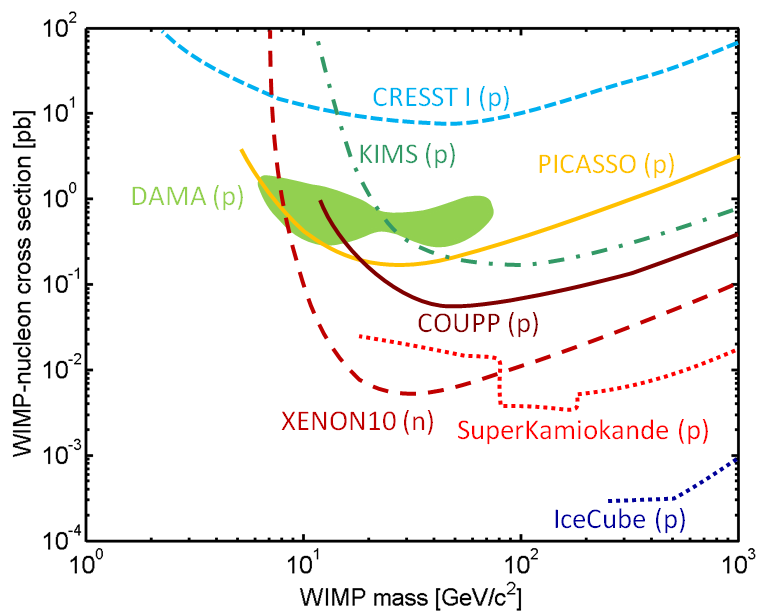}
\end{center}
\vspace*{-0.6cm}
\caption{\small \it Limits on the spin-dependent WIMP-nucleon cross section from some of the experiments discussed here. From top to bottom ordered by the minimum of the curves: CRESST I \cite{CRESST_I}, short-dashed; KIMS \cite{KIMS}, dash-dotted; PICASSO \cite{PICASSO}, light solid; COUPP \cite{COUPP_II} dark solid; XENON10 \cite{XENON10}, dashed; SuperKamiokande \cite{SK}, light dotted; IceCube \cite{IceCube}, dark dotted. Also shown is the 5$\sigma$ region allowed by DAMA as interpreted by Savage et al.\ \cite{Savage}, shaded region. Note that the XENON10 limit assumes a WIMP interactions with the neutron spin while all others assume a WIMP-proton spin interaction.}
\label{SD}
\end{figure}

Two of the experiments discussed here, DAMA and CoGeNT have reported evidence for a WIMP signal. While the annual modulation in the DAMA data is obvious, it is less clear how the CoGeNT data lead to a preferred cross section range given that the featureless low-energy part of the spectrum which is claimed to provide the evidence for WIMP interactions is well fit by the no-WIMP hypothesis \cite{CoGeNT}. Nevertheless, we can compare these claims with the results from other experiments. Figures \ref{SI} and \ref{SD} show the results for spin-independent and spin-dependent WIMP-nucleon interaction respectively from most of the experiments discussed here (some of the earlier results are left out and most are only reported in either the spin-dependent or the spin-independent plot). In addition to the limits from experiments which see no evidence for a WIMP signal, these plots show a preferred WIMP parameter region for the DAMA experiment as calculated by C. Savage et al.\ \cite{Savage}.

As can be seen, the spin-dependent interpretation of the DAMA signal is in strong tension with the null results of other experiments. The PICASSO data close the last previously still allowed window at low WIMP masses.

\begin{figure}[ht]
\begin{center}
\includegraphics[width=10cm]{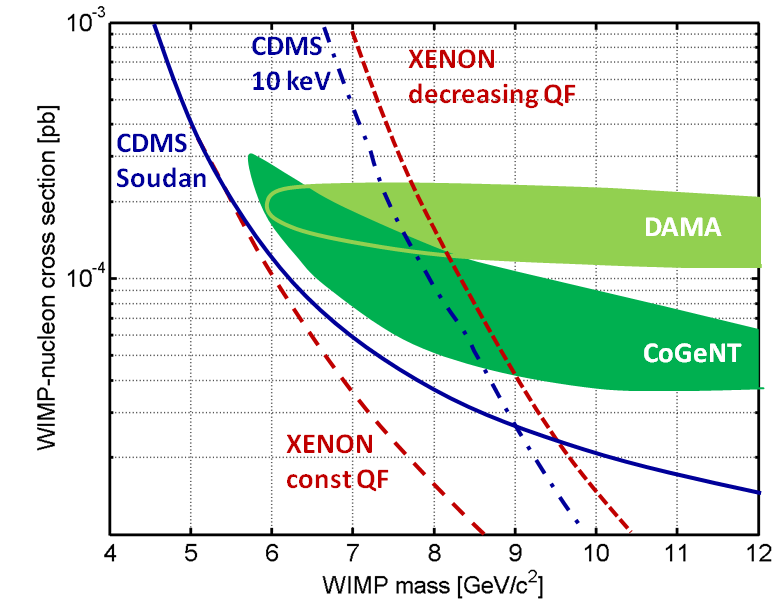}
\end{center}
\vspace*{-0.6cm}
\caption{\small \it Preferred WIMP parameter regions extracted from the DAMA data \cite{DAMA_I} as interpreted by \cite{Savage} (light shaded), and from CoGeNT \cite{CoGeNT} (dark shaded). The first interpretation of the XENON100 data (\cite{XENON100} lower dashed line) is incompatible with these evidence regions but this interpretation has been criticized due to systematic uncertainties regarding the quenching factor. More conservative assumptions lead to the upper short-dashed line \cite{CDMS_lowM_Soudan}. However, an analysis of the low energy region of CDMS data, previously disregarded due to less efficient background discrimination leads to an upper limit incompatible with the standard WIMP interpretation of both, DAMA and CoGeNT data (\cite{CDMS_lowM_Soudan}, solid line).}
\label{SI_lowM}
\end{figure}

The CoGeNT region and the DAMA region, both under the assumption of spin-independent interaction, are shown in figure \ref{SI_lowM}. The results from CDMS discussed so far only cover part of the region preferred by the two experiments. The interpretation of the XENON100 data as reported in \cite{XENON100} is incompatible with the preferred regions; however, this interpretation has been criticized since the low-mass limit strongly depends on the assumptions made regarding the quenching factor (called $L_{eff}$ in \cite{XENON100}) at low energy. Other assumptions have been proposed and while it is not obvious what assumption is most reasonable, it becomes clear that the most conservative assumptions move the XENON100 limit considerably above the CoGeNT evidence region at low WIMP masses. The CDMS collaboration has re-analyzed previous data sets, taking into account the low energy region where the electron recoil discrimination efficiency is considerably reduced. This leads to a significant number of background events, but since the recoil spectra for low-mass WIMPs rise very steeply at low energy a competitive limit can still be extracted. This new analysis which is less affected by systematic uncertainties than the XENON100 limit at low mass is incompatible with the standard WIMP interpretation of the DAMA and CoGeNT results.\cite{CDMS_lowM_Soudan}\cite{CDMS_lowM_SUF}

So far no convincing conventional interpretation has been proposed for the annual modulation signal observed by DAMA. Since the conventional WIMP interpretation is in strong tension with other experiments, alternative dark matter models have been explored. Inelastic dark matter models have been proposed, where the dark matter particle has a low lying excited state (several tens to a couple of hundred keV) and the elastic scattering process is highly suppressed. In this case only inelastic scattering is possible, but due to the necessary excitation energy this process is highly suppressed for lighter target nuclei, evading e.g.\ the tension between DAMA and CDMS (see e.g.\ \cite{InelDM}) However, this model seems to be incompatible with the results from CRESST which should show a significant signal due to the heavy tungsten nuclei. \cite{CRESST_Inel}

Other models, invoking electron recoil type interactions have been proposed, but so far none of those has found independent experimental support. While a dark matter interpretation of the DAMA modulation is not necessarily completely excluded, it becomes more and more difficult to come up with models that can explain the null-result of other experiments and at the same time are compatible with astrophysical observations.

\section{Conclusion}
Overwhelming evidence exists for the presence of large amounts of non-baryonic dark matter in the universe which forces us to extend the standard model of particle physics. WIMPs are prime candidate particles to solve the dark matter problem, but so far no convincing experimental evidence for such particles has been found in either indirect search for annihilation products or direct search for WIMP interactions with ordinary matter. However, the sensitivity of the experimental efforts is just starting to probe the parameter range preferred by theoretical models. Many collaborations are presently working on larger scale experiments with an expected improvement in sensitivity of one to two orders of magnitude. Large mass cryogenic detectors (EURECA in Europe and SuperCDMS and GEODM in the North America) and noble liquid detectors (e.g. XENON1T or DEAP360) are proposed or being prepared. With these efforts we have a realistic chance to find an answer to the question of what makes up more than 80\% of the matter in the Universe.

While dark matter searches may have a chance to detect new particles, their opportunities to study details of the underlying theory which describes those particles are very limited. Here we can expect help from a very different branch of particle physics: with the upcoming results from the new particle accelerator at CERN in Geneva, the Large Hadron Collider (LHC), we expect to learn more about possible extensions of the standard model of particle physics. Here we can hope to produce so far unknown particles, including WIMP candidates.\cite{LHC} However, accelerator experiments will not be able to prove that what they find comprises the dark matter in the universe since it will be e.g.\ impossible to demonstrate that a candidate particle does not decay on cosmological time scales.

With the great developments of the past years in direct and indirect dark matter search and the support from LHC we have a great set of tools to investigate the dark matter problem and I would not be surprised if in the not too distant future we have an answer, either can celebrate a discovery or at least know that we have to move on to different explanations.

\end{document}